\documentclass[english,aps,prper,reprint,showpacs,titlepage,longbibliography]{revtex4-2}   
\usepackage[utf8]{inputenc}
\usepackage[compatibility=false]{caption} % due to error note 
\usepackage{booktabs} % For better-looking tables
\usepackage{tabularx} % For flexible column widths
\usepackage{ragged2e} % For text alignment
\usepackage{caption} % For caption customization
\usepackage[T1]{fontenc}	
\usepackage{geometry}
\usepackage{times}
\usepackage{hyperref}  
\hypersetup{colorlinks=true,urlcolor=blue,citecolor=blue,linkcolor=blue} 
\usepackage{array}
\usepackage{enumerate}
\usepackage{amsmath}
\usepackage{amssymb}
\usepackage{tikz}
\usepackage{graphicx}
\usepackage{multirow}
\usepackage{tcolorbox}
\usepackage{ragged2e}
\usepackage{float}
\usepackage[utf8]{inputenc}

\usepackage[normalem]{ulem}
\usepackage{setspace} % Include the setspace package for line spacing control
\usepackage{graphicx}    % For including images
\usepackage{caption}     % For customizing captions
\usepackage{subcaption}  % For subfigure support (modern alternative to subfigure)

\begin{document}
\begin{titlepage}

\title{Students' Perceptions to a Large Language Model's Generated Feedback and Scores of Argumentation Essays}

 \author{Winter Allen}
 \affiliation{Department of Physics and Astronomy, Purdue University, 525 Northwestern Ave, West Lafayette, IN-47907, U.S.A.} 

 \author{Anand Shanker}
 \affiliation{Purdue University, 799 W Michigan St, Indianapolis, IN 46202}
  
 \author{N. Sanjay Rebello}
 \affiliation{Dept. of Physics and Astronomy / Dept. of Curriculum \& Instruction, Purdue University, West Lafayette, IN-47907, U.S.A.} 

\keywords{}

\begin{abstract}

% Problem-solving is a critical skill for success in science, engineering, and the broader workforce. However, students in introductory physics courses often rely on ineffective strategies, focusing on final answers rather than understanding underlying principles. Scientific argumentation has been shown to enhance students' problem-solving abilities and critical thinking. Within Physics Education Research (PER), integrating scientific argumentation into problem-solving fosters critical thinking and links conceptual knowledge with practical application. By facilitating learners to articulate their scientific arguments for solving problems, and by providing real-time feed-
% back on students’ strategies, we aim to enable students to develop
% superior problem-solving and metacognitive skills. Despite these benefits, providing timely, individualized feedback to students in large-enrollment physics courses remains a challenge. Recent advances in Artificial Intelligence (AI) offer promising solutions. This study investigates the potential of AI-generated feedback on students’ written scientific arguments in an introductory physics class. Using Open AI's GPT-4o, we provided delayed feedback on student written scientific arguments and surveyed them about the perceived usefulness and accuracy of this feedback. Our findings offer insights into the viability of implementing real-time AI feedback to enhance students’ problem-solving and metacognitive skills in large enrollment classrooms.

Students in introductory physics courses often rely on ineffective strategies, focusing on final answers rather than understanding underlying principles. Integrating scientific argumentation into problem-solving fosters critical thinking and links conceptual knowledge with practical application. By facilitating learners to articulate their scientific arguments for solving problems, and by providing real-time feedback on students’ strategies, we aim to enable students to develop superior problem-solving skills. Providing timely, individualized feedback to students in large-enrollment physics courses remains a challenge. Recent advances in Artificial Intelligence (AI) offer promising solutions. This study investigates the potential of AI-generated feedback on students’ written scientific arguments in an introductory physics class. Using Open AI's GPT-4o, we provided delayed feedback on student written scientific arguments and surveyed them about the perceived usefulness and accuracy of this feedback. Our findings offer insights into the viability of implementing real-time AI feedback to enhance students’ problem-solving and metacognitive skills in large-enrollment classrooms.

% Background/Context – Briefly introduce the problem or topic and its significance.
% Objective/Purpose – Clearly state the research question or aim.
% Methods/Approach – Summarize the methodology or framework used.
% Key Findings/Results – Highlight the most important outcomes or insights.
% Conclusion/Implications – Explain the impact, relevance, or next steps.
  
    \clearpage
\end{abstract}

\maketitle
\end{titlepage}
\maketitle

\section{Introduction}
Problem solving is a highly valued skill that is essential for participating in today's workforce. Learning to define problems and design solutions are key science and engineering practices identified in the Next Generation Science Standards (NGSS) \cite{NGSS2013}. A vast body of literature shows that students in introductory STEM courses often use ineffective problem-solving strategies such as means-ends analysis \cite{leonard1996using,maloney2011overview} without understanding the underlying principles, reflecting on them, or considering alternatives \cite{dufresne1997solving}. Students often prioritize memorizing final answers over developing a deeper understanding of the problem-solving process \cite{Tulminaro2007,dufresne1997solving}. To foster this growth, it is crucial not only to understand why students solve problems the way they do but also to help them reflect on their own problem-solving strategies, allowing them to develop as both learners and future scientists.

Scientific argumentation is a proven strategy to improve critical thinking that provides a schema for justifying the relevance of the retrieved knowledge in problem solving. To construct an argument students must justify their methods and decisions as they solved a problem, go through every step they took up to their solution, and provide evidence and reasoning for their process. In the context of problem-solving in physics, scientific argumentation involves not only an explanation of conceptual knowledge and methods but the ability to justify reasoning with empirical evidence and logical consistency. Within Physics Education Research (PER), scientific argumentation has been shown to enhance students' ability to link theoretical knowledge with practical problem-solving skills \cite{rebello2019using}. This process encourages students to think critically about the methods they use and the evidence they gather, promoting skills that are essential for expert-like problem solving. The iterative nature of scientific argumentation aligns with the goals of PER in promoting both content mastery and the development of scientific critical thinking.

By facilitating learners to articulate their scientific arguments for solving problems, and by providing real-time feedback on students’ strategies, we aim to enable them to develop superior problem-solving and metacognitive skills. However, we face the daunting task of giving students timely, relevant feedback on their problem solving. In large enrollment courses, the time and effort needed to provide individualized feedback on students' strategy essays is prohibitive. This problem has existed for many years; however, the recent advancements in Artificial Intelligence (AI) may offer a solution \cite{siverling2021initiates}. Recently, Large Language Models (LLMs) have undergone major developments. Many researchers in PER are exploring the use of LLMs, such as GPT-4, to explore the grading of student open responses \cite{PhysRevPhysEducRes.20.020144,henkel2024largelanguagemodelsmake,chen}.

To explore the need and usefulness of AI generated feedback for our students' written scientific arguments, we designed a study to provide students with delayed feedback to their written arguments utilizing OpenAI's GPT-4o. This study was conducted on a quiz bonus question in an introductory physics class to determine the viability of eventually implementing real-time feedback to students in our large enrollment class. As with any classroom, our main concern is always the benefit of students. Our research questions are: 
\\

%\textit{To what extent can an LLM score and provide feedback in a way that students find useful and accurate?}
\textit{(1) How does the score provided by an LLM on a student strategy essay compare for essays written by students who answered the question correctly versus those who answered it incorrectly? (2) What are students' perceptions about the usefulness and accuracy of the LLM feedback?}
\\

%We report on the LLM's scoring as well as students' perceived usefulness and accuracy of the LLM-generated feedback collected from an extra credit post-survey.
\section{Background} 

Scientific argumentation has been identified as a key science and engineering practice specified in the NGSS \cite{NGSS2013}. Research suggests that students tend to struggle with the idea of developing scientific arguments\cite{berland2009making, kuhn1993science}, especially with finding appropriate evidence and constructing their reasoning \cite{berland2010learning,kuhn2010teaching} and distinguishing between various elements of an argument \cite{forman1998you,jimenez2000doing}. To aid students in constructing arguments,argumentation scaffolds can elicit students’ participation in scientific argumentation \cite{jonassen2010arguing}. Appropriate scaffolds include justification prompts \cite{xun2004conceptual} and question prompts \cite{cho2002effects} in instructional materials that help students articulate the rationale for their problem-solving steps and urge them to reason using evidence and justifications \cite{christodoulou2014science,mcneill2010scientific} based on underlying principles \cite{schworm2007learning}. However, most undergraduate physics courses do not facilitate argumentation. Curricula that facilitate more expert-like problem solving can positively influence students’ epistemic beliefs and expectations about problem solving \cite{wampler2013relationship}. In more recent work, Rebello et al. \cite{rebello2019scaffolding,rebello2019using} found positive effects of using scientific argumentation in physics courses for future elementary teachers as well as future engineers. 

McNeill and Krajcik \cite{mcneill2011supporting} adapted Toulmin's \cite{toulmin2003uses} argumentation protocol to a scientific argument as comprising three elements: claim, evidence, and reasoning (CER). In this model, the claim is an assertion or conclusion about a phenomenon, the evidence consists of scientific data supporting the claim, and the reasoning explains the relevance of the evidence. CER has become a popular framework in K-12 education, where students are encouraged to construct arguments using data to support their claims \cite{mcneill2008inquiry, wang2020scrutinising}. Given the effectiveness of CER in K-12, there is a strong rationale for exploring its adaptation in undergraduate physics education, where developing students' ability to argue scientifically can enhance their problem-solving and critical thinking skills. In order for students to improve their argumentation, prompt feedback is an important tool.

Feedback can be defined as information regarding aspects of a learner’s performance or understanding \cite{hattie2007power}. Research \cite{burgess2020feedback} has shown that feedback is one of the important drivers of learning. Feedback can facilitate improvements in learners’ understanding and skills \cite{henderson2021usefulness,nicol2006formative, sadler2010beyond,shepard2000role} by informing learners about their progress, reinforcing good practice, and motivating them to engage in self-regulation \cite{burgess2015receiving,shepard2000role}. It facilitates self-assessment and reflection on performance, which can narrow the gap between actual and desired performance \cite{burgess2020feedback}. 

Research  \cite{wisniewski2020power} has shown that the effectiveness of feedback increases with the information that it contains. Previous research \cite{kluger1996effects} has also shown that moderators such as timing, specificity, and task complexity affect how learners receive and use feedback \cite{hattie2018visible,brooks2019matrix}. Feedback is most effective when it is integrated into the learning process through formative assessment \cite{branch2002feedback,hattie2007power} and provided prior to completion \cite{henderson2021usefulness}. It is also effective if it is sufficiently detailed \cite{ferris1997influence,price2010feedback}, usable \cite{hepplestone2014understanding, Winstone02012017}, and facilitates change \cite{ryan2019feedback}, such that learners can test their new understandings \cite{boud2013feedback,court2014tutor,Pitt19052017}. In asynchronous and isolated online settings \cite{orlando2011effectively}, interactive dialogues can be especially useful \cite{wolsey2008e} as students cannot easily interact with their peers \cite{Furnborough01112009} which puts significant weight on the feedback comments they receive \cite{ortiz2005college}. 

The effectiveness of feedback is also influenced by its cognitive complexity. Task Level feedback focuses on completion and correctness of the task. Process Level feedback focuses on the strategies used by the learners to understand and master the tasks. Finally, Self-Regulation Level feedback focuses on helping the learner manage, guide, and monitor their own learning and actions. Most effective feedback includes information at all three levels as it helps learners not just understand what mistakes they made, but also the underlying reasons and strategies to avoid them in the future \cite{hattie2007power}. 

By combining the underlying extensive knowledge of beneficial feedback with modern tools, such as LLMs, there is potential for providing students with constructive feedback that aids in their learning, especially in a large enrollment course where students would normally not be able to receive individualized feedback.

\begin{table*}[htbp]
\caption{Average scores  of written arguments given by the LLM of students who got the multiple-choice correct vs. incorrect}
\begin{ruledtabular}
\begin{tabular}{cccccc} 
 & \textbf{Total Average $\pm$ SD}  & \textbf{Correct Average $\pm$ SD}  & \textbf{Incorrect Average $\pm$ SD}  & \textbf{p-value} & \textbf{Cohen's d}\\ 
\hline
\\
\textbf{Quiz 08} & 1.69 $\pm$ 1.29 & 2.14 $\pm$ 1.24 & 1.12 $\pm$ 1.11 & $<< 0.001$ & 0.87 \\
\\
\textbf{Quiz 09}  & 2.72 $\pm$ 1.08 & 2.77 $\pm$ 1.08 & 2.53 $\pm$ 1.04 & 0.015 & 0.22 \\

\end{tabular}
\label{LLMscores}
\end{ruledtabular}
\end{table*}

\begin{figure*}[t]
    \centering
    \begin{subfigure}[t]{0.45\linewidth}
        \includegraphics[width=\linewidth]{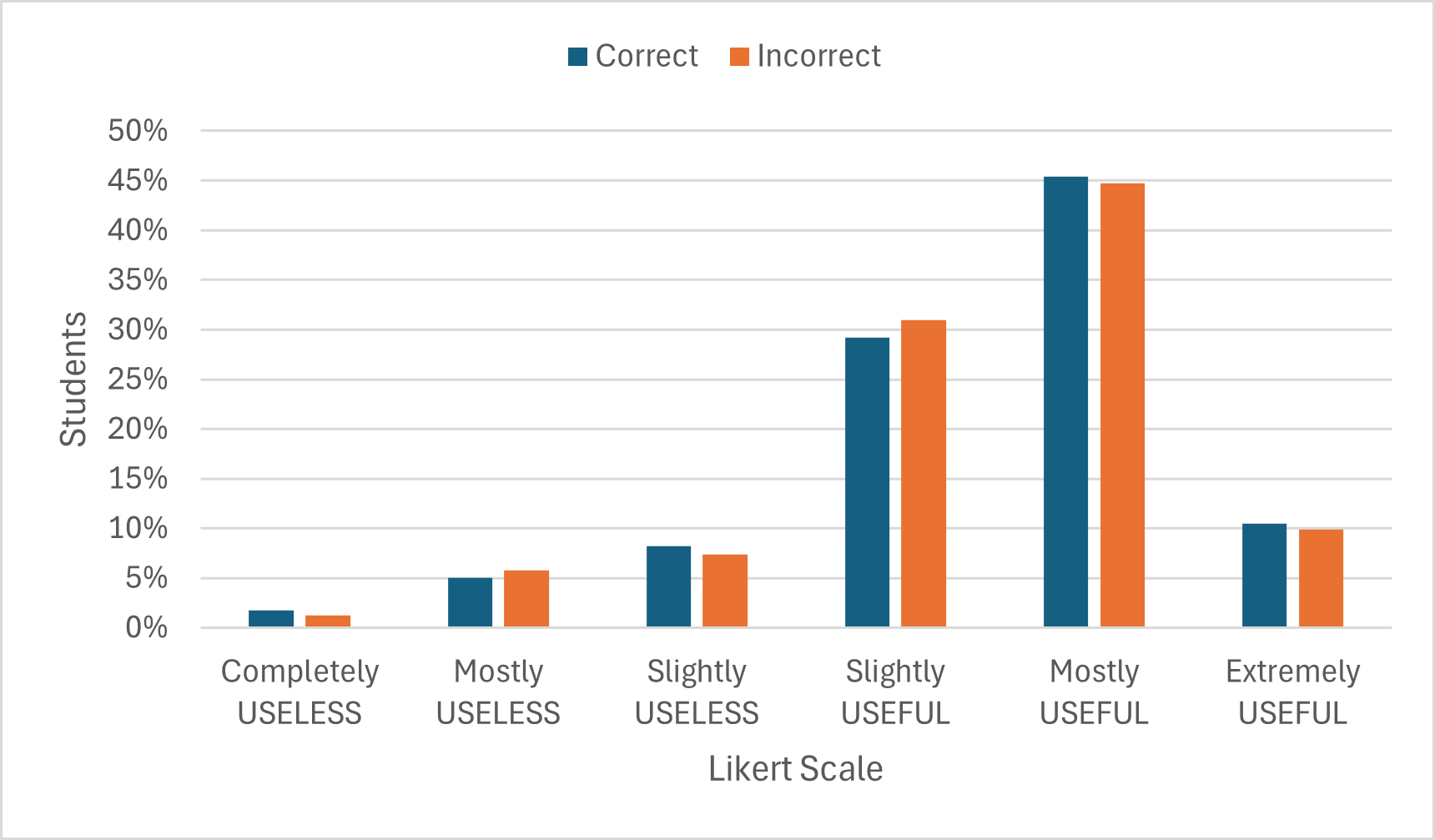}
        \caption{Quiz 08 Perceived Usefulness}
        \label{fig:q8_usefulness}
    \end{subfigure}
    \hfill
    \begin{subfigure}[t]{0.45\linewidth}
        \includegraphics[width=\linewidth]{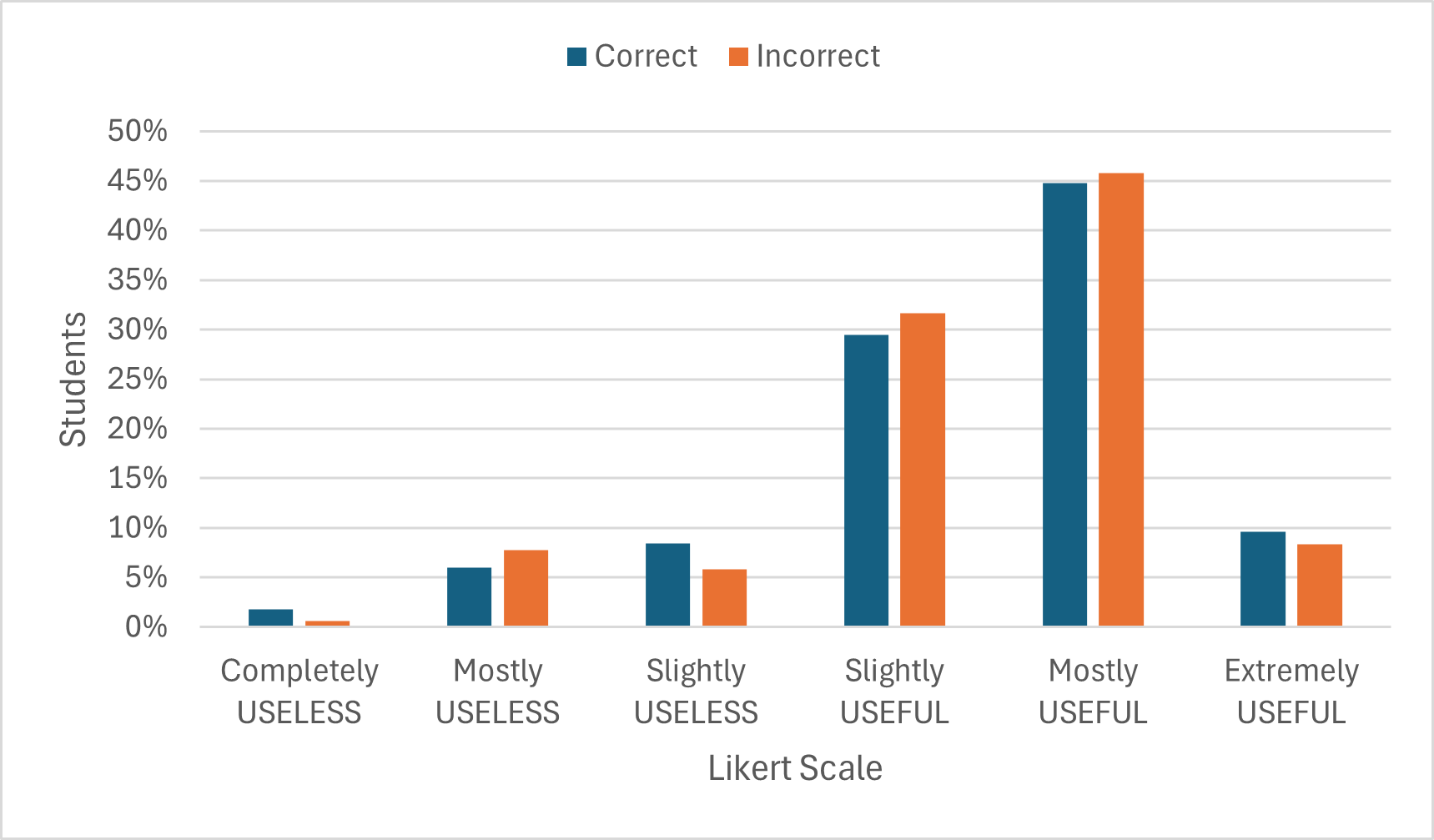}
        \caption{Quiz 09 Perceived Usefulness}
        \label{fig:q10_usefulness}
    \end{subfigure}
    \hfill
    \begin{subfigure}[t]{0.45\linewidth}
        \includegraphics[width=\linewidth]{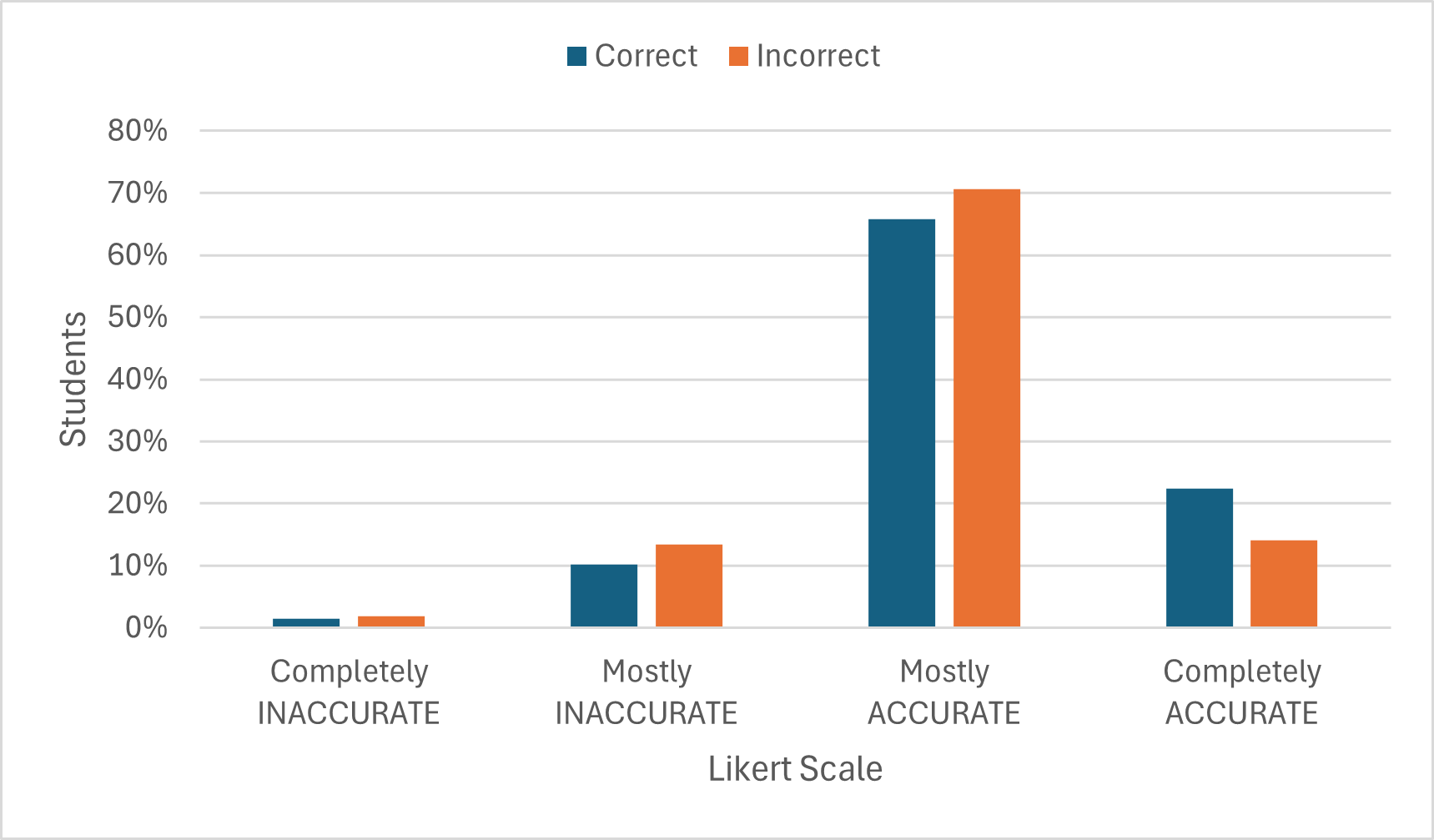}
        \caption{Quiz 08 Perceived Accuracy}
        \label{fig:q8_accuracy}
    \end{subfigure}
    \hfill
    \begin{subfigure}[t]{0.45\linewidth}
        \includegraphics[width=\linewidth]{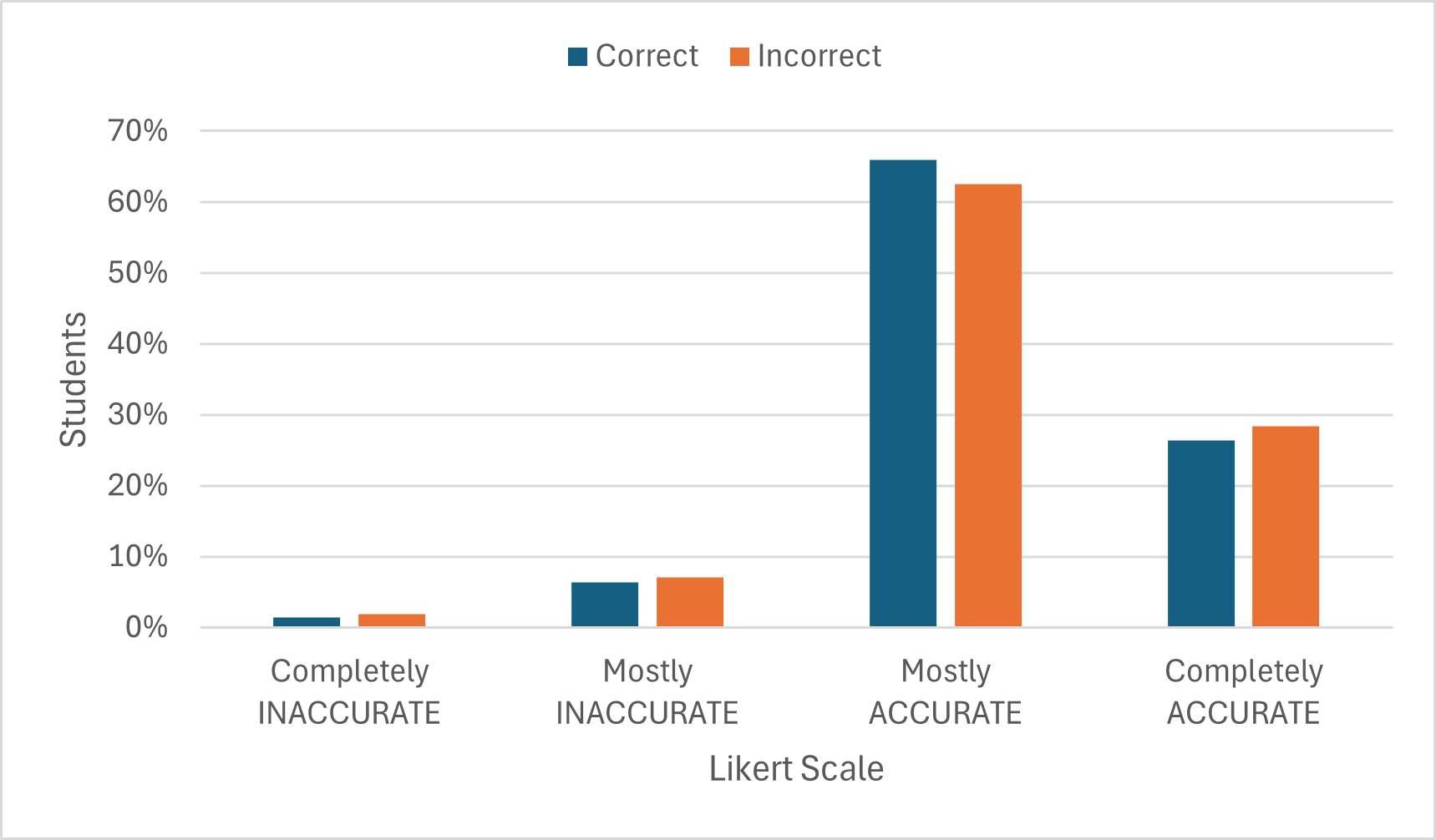}
        \caption{Quiz 09 Perceived Accuracy}
        \label{fig:q10_accuracy}
    \end{subfigure}
    \caption{Students Perception of the LLM Feedback.}
    \label{fig:perception}
\end{figure*}
\section{Methods}
\subsection{Context}
This study was conducted in a first semester calculus-based physics course, primarily for future engineers, at a large U.S Midwestern land-grant university. The course is sectioned into three modules, each focusing on a key physics principle. Students were tasked with writing a scientific argument to support their problem solving process for a multiple-choice question on an online quiz in the last five weeks of the semester. The students were taught scientific argumentation based upon claims, evidence, and reasoning (CER) \cite{mcneill2011claims}  through a series of scaffolds implemented in the recitation portion of the course. By the last 5 weeks, students were expected to be able to fully construct arguments based upon CER. 

To facilitate students' argumentation skills, various levels of scaffolding were provided as training for the students throughout the semester. In Weeks 01-05 of the semester, students received various statements at the end of the recitation file. Their goal was to assign each statement the label of claim, evidence, or reasoning. After the submission deadline, the correct labels would be released in the solution document of the recitation. In Weeks 06-10 of the semester, students were asked to write their own CER statements. They would provide at least one statement for claims, evidence, and reasoning. In Weeks 11-15 of the semester students were expected to construct full and complete arguments with minimal scaffolding. The full process is outlined in detail by Allen et al.\cite{allen2025assessing}. 

% The argumentation prompt that students were provided at the end of the recitation notebook was:
% \\
% \textit{A scientific argument consists of three pieces: CLAIM, EVIDENCE, and REASONING.
% \\A CLAIM is every decision you took to solve the problem i.e. the choice of principle / concept, system / surroundings, approximation / assumption, or key steps you used in your problem-solving strategy – It is an assertion that that the decision was the right one to answer the key question(s) in the problem statement.
% \\EVIDENCE is the information either provided in the problem or something that you know (e.g. from the physics class) to be factually correct that helped you make the decision above.
% \\REASONING is an explanation of why the EVIDENCE supports the CLAIM.
% In words, construct an argument to explain, elaborate and justify your solution for Part 2. Your argument should be in a paragraph and contain the CLAIMS, EVIDENCE, and REASONING that support your solution.}
% \\

This study focuses on arguments students constructed to the first question on quizzes 08 and 09 in the last module of the course. The quizzes were administered online through an anti-cheating Lockdown Browser. For both quizzes, the problem centered around energy. Once students solved the problem, they were prompted to write an argument: 
\\

\textit{Describe using WORDS ONLY, a scientific argument for how you solved the previous problem. Do not use any numbers, symbols, or formulae in your answer. Your response should be at least 50 words long.}
\\
\subsection{Feedback Generation}
Student responses were downloaded from our Learning Management System (LMS) and given to OpenAI's GPT 4o \cite{openai_gpt4o} using its application programming interface (API). For both quizzes, the LLM was given the prompt: \textit{You are an educator. Your goal is to provide useful feedback on the physics aspect of the essay. Your feedback should be no more than 100 words long. Focus only on the physics ideas and concepts. Do not include salutations. Do not comment on the grammar and sentence structure. Do not include complements or critiques.} In addition, the temperature was set to 0.80, and we provided the problem students solved along with discouraging the LLM from focusing on aspects that did not focus on the physics such as salutations, complements, and grammar. The API was fed an example of an ideal answer, along with a rubric. The rubric was out of 5 points. The distribution allotment of points differed slightly between quizzes; however, 2 to 3 points were given for accurately identifying the claims that center around the appropriate physics principle with the other 2 to 3 points focusing on appropriate evidence. The rubrics were designed to mirror the scaffolding students received in the Recitation portion of the course.

\subsection{Data Collection}
To determine how students felt about the feedback from the LLM, they were given bonus points if they filled out a delayed post-survey approximately 1-2 weeks after they wrote their arguments. Students were given the feedback via a tiny URL. Once students viewed their written argument, the score from the LLM, and the feedback from the LLM, they were asked four questions. One was a six-point Likert scale question asking how useful students found the feedback from "completely useless" to "extremely useful". Another was a four-point Likert scale question of how accurate students found the feedback from "completely inaccurate" to "completely accurate". The last two questions focused on students' comments and suggestions for improvement of the feedback. For this study, we will focus on the Likert scale questions, while the open ended questions will be analyzed in a future work.  The survey and student data were administered and extracted through our course LMS, respectively. 

We report on the distribution of ratings given by students for the LLM's usefulness and accuracy. We also report on the average of scores the LLM gave to students who answered the multiple-choice question correct versus incorrect. These results will give us an idea of how students perceive the LLM and if there is a statistically significant difference of average argument scores between the student who answered the multiple-choice question correct versus incorrect.

\section{Results \&\ Discussion}

To assess how the LLM scored students, we first calculated the average of LLM generated scores compared for students who got the multiple-choice (MC) question that the argumentation was based upon correct or incorrect. These are reported in Table \ref{LLMscores}. 

In Quiz 08, students (N=730) solved a high difficulty problem, as the average to the MC problem was about ~47\%\, and the average LLM generated score was 1.69 out of 5 points which is about ~34\%\ .  The  students who answered the MC question correctly were scored on average at 2.14 (~43\%) and those who answered it incorrectly were scored on average 1.12 (~22\%). To determine whether the difference between the average scores of correct and incorrect students was statistically significant, we did a t-test and measured the effect size. Due to the very small p-value and large Cohen's d, we report that the average scores are statistically significant. This can be interpreted as the LLM scores of the written arguments of students who selected the correct MC answer differed from the LLM scores of students who selected the incorrect MC answer on a high difficulty problem. 

In Quiz 09, students (N=565) solved a low difficulty problem, as the average on the MC problem was about ~76\%\, and the average LLM generated score was 2.72 out of 5 points which is about ~54\%\ . The students who answered the MC question correctly were scored on average at 2.77 ( ~55\%\ ) and those who answered it incorrectly were scored on average 2.53 ( ~51\%\ ). To determine whether the difference in average scores between correct and incorrect students was statistically significant, we did a t-test and measured the effect size. A very small p-value and large Cohen's d indicate that the average scores are statistically significant but the impact is small. It is important to note that the average for this quiz question was significantly higher than that of Quiz 08, implying students had less difficulty solving the problem. Overall, we find that, for Quizzes 08 and 09, LLM scored students as expected, given quiz averages, which indicates that the LLM scoring works for both high and low difficulty problems.

After reviewing their feedback and scores from both quizzes, students were asked to rate the usefulness and accuracy of the feedback on a six-point Likert scale from "completely useless" to "extremely useful" and "completely inaccurate" to "extremely accurate". The results are shown in Figure \ref{fig:perception}. The orange bars represent students who got the MC question incorrect, and the blue bars represent students who got it correct. Across both quizzes, where the average scores differed significantly, the majority of students found the feedback from the LLM to be "mostly useful" and "mostly accurate". There is no significant difference between how students who got the MC question incorrect versus correct scored the feedback. This is an extremely promising result as it indicates that students found the LLM's feedback on their essay to be useful whether or not they answered the MC question correctly on both high and low difficulty problems, indicating that most students perceive LLM feedback as both useful and reliable.

An example of feedback that one student found extremely useful and mostly accurate where they received a score of 3 out of 5 is :

\textit{The student's essay effectively mentions the extended system energy principle, and correctly describes the change in translational kinetic energy as equal to the net work by external forces. It also captures the idea that the total work done on the system equals the change in translational kinetic energy plus the change in internal energy. However, it does not explicitly mention the point particle energy principle, nor does it mention any of the assumptions regarding speed, force, friction, or air drag.} 

This example illustrates that the LLM focuses on the key physics without delving into the nuances of language as much. By exploring more examples of LLM generated feedback and the written feedback of students in a future work, we will likely be able to gain a deeper understanding of feedback that is most beneficial to student growth.

\section{Conclusions, Limitations and Future Work}

The ability to provide students with prompt and constructive feedback using the recent developments in LLMs would be invaluable to large enrollment courses where providing individualized feedback is prohibitive. The major takeaway of this exploratory study is that utilizing LLMs to provide feedback to students' written arguments may be a viable option. Our question centered around whether students find LLM generated feedback beneficial. From our survey, we found that students considered the feedback mostly useful and accurate. By continuing to explore this area and develop the prompting and feedback, there could be important implications in student learning. Students who previously would have received little to no feedback on their written essays will eventually be able to receive individualized and constructive real-time feedback in their studies.

There are a few limitations to this study. Firstly, the feedback was delayed. Students received feedback only at the end of the semester and only for three quizzes. As a result, there was likely little benefit to the progression of student written argumentation. To address this, we intend to provide feedback more promptly, starting with shortly after quizzes are completed to eventually real-time feedback. The goal is to improve student argumentation throughout the semester.

We also acknowledge that different methods of prompting can be explored. This can go as far as prompting the LLM with specific guidelines for administering feedback, based on prior feedback research. In the immediate future, we will analyze student written responses and suggestions to improve the feedback from the LLM, along with comparing the LLM-generated scores and feedback to a human grader's scores and feedback. Using this information, we will explore how different prompting methods can improve LLM feedback for students. This will hopefully lend itself to adapt the feedback in ways students find more beneficial to their growth and learning.  

\section{Acknowledgments}
This work is supported in part by U.S. National Science Foundation grants 2300645 and 2111138. Opinions expressed are of the authors and not of the Foundation

\clearpage
\bibliography{references}

\end{document}